\documentclass[a4paper,showpacs,aps,prd,nofootinbib]{revtex4}
\usepackage[latin9]{inputenc}
\usepackage{hyperref}
\hypersetup{
  colorlinks=true,        
  linkcolor=blue,         
  citecolor=cyan,         
}
\usepackage{breakurl}
\usepackage{graphicx}
\usepackage{epsf}
\usepackage{epsfig}
\usepackage{amssymb,amsmath}
\usepackage[usenames]{color}
\usepackage{amssymb}
\usepackage{times}



\begin{document}

\title{Perturbation solutions of relativistic viscous hydrodynamics for longitudinally expanding fireballs}

\newcommand{\orcidauthorA}{0000-0002-2193-6487}

\author{Ze-Fang Jiang$^{~1,2}$}
\email{jiangzf@mails.ccnu.edu.cn}
\author{Duan She$^{~2,3}$}
\email{sheduan@mails.ccnu.edu.cn}
\author{C. B. Yang$^{~2}$}
\email{cbyang@mail.ccnu.edu.cn}
\author{Defu Hou$^{~2}$}
\email{houdf@mail.ccnu.edu.cn}

\affiliation{$^1$ Department of Physics and Electronic-Information Engineering, Hubei Engineering University, Xiaogan 432000, China}
\affiliation{$^2$ Institute of Particle Physics \& Key Laboratory of Quark and Lepton Physics, MOE, Central China Normal University, Wuhan 430079, China}
\affiliation{$^3$ Physics Department and Center for Exploration of Energy and Matter, Indiana University, 2401 N Milo B. Sampson Lane, Bloomington, IN 47408, USA.}

\begin{abstract}
The solutions of relativistic viscous hydrodynamics for longitudinal expanding fireballs is investigated with the Navier-Stokes theory and Israel-Stewart theory.
The energy and Euler conservation equations for the viscous fluid are derived in Rindler coordinates with the longitudinal expansion effect is small.
Under the perturbation assumption, an analytical perturbation solution for the Navier-Stokes approximation and numerical solutions for the Israel-Stewart approximation are presented. The temperature evolution with both shear viscous effect and longitudinal acceleration effect in the longitudinal expanding framework are presented and specifically temperature profile shows symmetry Gaussian shape in the Rindler coordinates.
In addition, in the presence of the longitudinal acceleration expanding effect,
the results of the Israel-Stewart approximation are compared to the results from the Bjorken and the Navier-Stokes approximation,
and it gives a good description than the Navier-Stokes theories results at the early stages of evolution.
\end{abstract}
\maketitle
\date{\today}

\section{Introduction}
The Relativistic hydrodynamic theory provides well description of the space-time evolution and many non-equilibrium properties of quark-gluon plasma (QGP) produced
in heavy ion collisions at the Relativistic Heavy Ion Collider (RHIC) and the Large Hadron Collider (LHC)~\cite{Bass1998vz,Gyulassy:2004zy,Shuryak:2003xe,Heinz2013th}.

There has been a lot of excellent progress in solving relativistic viscous hydrodynamics equations analytically with different approximations and special symmetries and numerically in recent decades years~\cite{Romatschke:2017ejr,Israel:1979wp,AM:2004prc,Koide:2006ef,PeraltaRamos:2009kg,
Denicol:2012cn,Landau:1953gs,Hwa:1974gn,Bjorken:1982qr,Biro:2000nj,Csorgo:2003rt,Csorgo:2006ax,Borshch,Csorgo:2008prc,
Nagy:2009eq,Csanad:2012hr,Csorgo:2018pxh,Gubser:2010ze,Gubser:2010ui,Jiang2017,Jiang:2014uya,Jiang:2014wza,Hatta:2014gqa,Hatta:2014gga,
Wu:2016pmx,She:2019wdt,Schenke:2010rr,Giacalone:2017dud,Pang:2018zzo,Chen:2017zte,Wu:2018cpc,Calzetta:2019dfr}.
Those analytical solutions play a very important role in understanding the evolution dynamics and are good testbeds for numerical solutions.

Recently,
a series of interesting analytical solutions for longitudinally expanding relativistic perfect fluid were found by Budapest and Wuhan group~\cite{Csorgo:2006ax,Csorgo:2008prc,Jiang2017,Csorgo:2018pxh,She:2019wdt}.
These ideal hydrodynamics solutions combined with Buda-Lund model~\cite{Csorgo:1995pr} have been utilized for simulating QGP medium dynamic evolution
and readily reproduce the observed final state multiplicity distribution and its dependence on beam energy,
collision system, particle mass and freeze-out temperature~\cite{She:2019wdt,Biro:2000nj,Csorgo:2006ax,Nagy:2009eq,Jiang2017,Jiang:2018qxd,Kasza:2018qah}.

However, a lot of comparisons between experimental data and viscous hydrodynamic simulations found that the picture of QGP is a nearly perfect fluid but contains a small specific shear viscosity.
The shear viscosity ratio of QGP is very close to the lower bound $1/4\pi$ computed for $\mathcal{N}=4$ super-Yang-Mills (SYM) theory in the AdS/CFT correspondence~\cite{Policastro:2001yc,Baier:2007ix,Bhattacharyya:2008jc,Arnold:2011ja}.
In this paper, we will go beyond both the Cs\"org\H{o}-Nagy-Csan\'ad (CNC) solutions and the Cs\"org\H{o}-Kasza-Csan\'ad-Jiang (CKCJ) solutions of the relativistic perfect fluid for longitudinally expanding fireballs~\cite{Csorgo:2006ax,Csorgo:2008prc,Csorgo:2018pxh}
and present a perturbation analytical solution of the longitudinally expanding first order (Navier-Stokes limit) viscous hydrodynamic equations.
We furthermore present the numerical results of the second-order (Israel-Stewart limit) viscous hydrodynamics equations
as a piece of the longitudinally expanding fireballs theory based on assuming the relaxation time is small~\cite{Paquet:2019npk}.
We find that small shear pressure tensor relaxation time $\tau_{\pi}$ approximation solves
the unstable problem of the first order approximation, indicating the stability of the second order numerical results.
This study providing us a self-consistent first-order and second-order viscous hydrodynamic with longitudinal expanding dynamics, and
lead us to a better understanding of the relationship between viscosity effect and longitudinal acceleration effect
for the medium evolution in future phenomenological studies.

The organization of the paper is as follows. In Sec.~\ref{section2},
the 2nd viscous hydrodynamic equations are reconstructed in Rindler coordinates according to the Landau-Lifshitz formalism~\cite{Landau:1953gs}, and perturbation solutions are presented. In Sec.~\ref{section3}, numerical results of viscous hydrodynamics for longitudinal expanding fireball are investigated.
Brief summary and discussion are given in Sec.~\ref{section4}.

\section{The perturbation solutions to the longitudinally expanding flow}
\label{section2}
We work in the so-called Rindler-coordinates for which $\tau=\sqrt{t^{2}-r^{2}}$ is the proper time and $\eta_{s}=0.5\ln[(1+r/t)/(1-r/t)]$ is the space-time rapidity, where $x^\mu=(t,r_1,\dots,r_d)$ and $r=\sqrt{\Sigma_i r_i^2}$~\cite{Csorgo:2006ax,Csorgo:2008prc,Csorgo:2018pxh}.
We consider (1+1) dimensional fluid flow in (1+3) dimensions space-time since we focus on the perturbation solutions of a longitudinal expanding fireball with shear viscosity.
The flow 4-velocity field $u^{\mu}$ in the Cartesian coordinates (the Minkowski flat space-time) for this system is
\begin{eqnarray}
u^{\mu}=(\cosh\Omega, 0, 0, \sinh\Omega),
\label{e:velocity}
\end{eqnarray}
where flow rapidity $\Omega$ is a function of space-time rapidity $\eta_{s}$ and is independent of proper time $\tau$~\cite{Csorgo:2008prc}, with the 4-velocity normalized as $u^{\mu}u_{\mu}=1$.
The second-order hydrodynamic equations without external currents are simply given by
\begin{eqnarray}
    \partial_{\mu} T^{\mu\nu} &=&  0,
\label{e:Tmunu}
\end{eqnarray}
with the energy-momentum tensor $T^{\mu\nu}=\varepsilon u^{\mu}u^{\nu}
-p \Delta^{\mu\nu} + \pi^{\mu\nu}$, where $\varepsilon$ is the
energy density, $p$ the pressure, $g_{\mu\nu}$=$\mathbf{diag}$$(1,-1,-1,-1)$ the metric tensor,
and $\Delta^{\mu\nu}=g^{\mu\nu}-u^{\mu}u^{\nu}$ the projection operator which is orthogonal to the fluid velocity.
The shear pressure tensor $\pi^{\mu\nu}$ represents the deviation from ideal hydrodynamics and local equilibrium,
and it satisfies $u^{\mu}\pi_{\mu\nu}=0$ and is traceless $\pi^{\mu}_{~\mu}=0$ in the Landau Frame.

The energy density and pressure are related to each other by the equation of state (EoS),
\begin{equation}
\begin{aligned}
\varepsilon=\kappa p,
\label{e:eos}
\end{aligned}
\end{equation}
where $\kappa$ is usually related to the local temperature~\cite{Paquet:2019npk},
in this case we assume $\kappa$ to be a constant and independent of the temperature.

The fundamental equations of viscous fluid dynamics are established by projecting appropriately
the conservation equations of the energy momentum tensor Eq.~(\ref{e:Tmunu}). The conservation equations can be rewritten as,
\begin{equation}
\begin{aligned}
D\varepsilon=-(\varepsilon+p)\theta+\sigma_{\mu\nu}\pi^{\mu\nu},
\label{e:en1}
\end{aligned}
\end{equation}
and
\begin{equation}
\begin{aligned}
(\varepsilon+p)Du^{\alpha}=\nabla^{\alpha}p+\pi^{\alpha\mu}Du_{\mu}-\Delta^{\alpha\nu}\nabla^{\mu}\pi_{\mu\nu},
\label{e:eu1}
\end{aligned}
\end{equation}
respectively, where $D=u^{\mu}\partial_{\mu}$ is the comoving derivative and $\theta=\partial_{\mu}u^{\mu}$ is the expansion rate.

In terms of the 14-moment approximation result from~\cite{AM:2004prc,DTeaney}, $\partial_{\mu}s^{\mu}\geq 0$ reduce the corresponding thermodynamic forces.
The general traceless shear tensor $\pi^{\mu\nu}$ is~\cite{AM:2004prc,Baier:2007ix},
\begin{eqnarray}
\pi^{\mu\nu} &=&2\eta\sigma^{\mu\nu}-\tau_\pi \left[ \Delta^\mu_\alpha\Delta^\nu_\beta u^\lambda \nabla_\lambda \pi^{\alpha\beta} +\frac{1}{3}\pi^{\mu\nu} \theta\right] \notag- \lambda_1 \pi^{\langle\mu}_{\ \ \lambda} \pi^{\nu\rangle\lambda}-\lambda_2 \pi^{\langle \mu}_{\ \ \lambda} \Omega^{\nu\rangle\lambda}- \lambda_3 \Omega^{\langle \mu}_{\ \  \lambda}\Omega^{\nu\rangle \lambda},
\label{eq:pimunu}
\end{eqnarray}
with the symmetric shear tensor $\sigma^{\mu\nu}$ and the antisymmetric vorticity tensor $\Omega^{\mu\nu}$ defined as
\begin{eqnarray}
&\sigma^{\mu\nu}&\equiv\left(\frac{1}{2}(\Delta^{\mu}_{\alpha}\Delta^{\nu}_{\beta}+\Delta^{\mu}_{\beta}\Delta^{\nu}_{\alpha})
-\frac{1}{3}\Delta^{\mu\nu}\Delta_{\alpha\beta}\right)\partial^{\alpha}u^{\beta}, \label{eq:shear} \\
&\Omega^{\mu\nu}&\equiv \frac{1}{2}\Delta^{\mu\alpha}\Delta^{\nu\beta}(\nabla_\alpha u_\beta -\nabla_\beta u_\alpha),
\end{eqnarray}
where $\eta$, $\tau_{\pi}$, $\lambda_{1}$, $\lambda_{2}$, $\lambda_{3}$ are positive transport coefficients in the flat space time. $\eta$ is the shear viscosity coefficient and $\tau_{\pi}$ is the relaxation time for shear pressure tensor corresponding to the dissipative currents, respectively.
Shear viscosity ratio $\eta/s$ of the QGP is very close to the lower bound $1/4\pi$ computed for a strongly coupled gauge theory
($\mathcal{N}=4$ SYM) in the AdS/CFT correspondence.
And relaxation time $\tau_{\pi}$ is in fact approximately $(2-\ln 2)/(2\pi T)$~\cite{Policastro:2001yc,Baier:2007ix,Bhattacharyya:2008jc,Arnold:2011ja}, where $s$ is the entropy, $s^{\mu}$ the entropy four-current, $T$ the temperature. It is customary to split $\pi^{\mu\nu}$ order-by-order in terms of $\sigma^{\mu\nu}$ into a traceless part and the contribution form higher-order term is suppressed by the relaxation time $\tau_{\pi}$, and assuming transport coefficients $\lambda_{1}=\lambda_{2}=\lambda_{3}=0$~\cite{AM:2004prc,DTeaney}, and neglect the contribution from higher order $\tau_{\pi}$ terms, one shows that Eqs.~(\ref{e:en1}-\ref{e:eu1}) can be cast into,
\begin{equation}
\begin{aligned}
D\varepsilon&=-(\varepsilon+p)\theta+2\eta \sigma^{\mu\nu}\sigma_{\mu\nu}-2 \eta\tau_{\pi}\sigma_{\mu\nu}\left[\Delta_{\alpha}^{\mu}\Delta_{\beta}^{\nu} D \sigma^{\alpha\beta} + \frac{1}{3}\sigma^{\mu\nu} \theta \right],
\label{en1}
\end{aligned}
\end{equation}
and
\begin{equation}
\begin{aligned}
(\varepsilon+p)Du^{\alpha}&=\nabla^{\alpha}p+2\eta\left(\sigma^{\alpha\mu}-\tau_{\pi}\left[\Delta_{\gamma}^{\alpha}\Delta_{\rho}^{\mu} D \sigma^{\gamma\rho} + \frac{1}{3}\sigma^{\alpha\mu} \theta \right]\right)Du_{\mu}\\
&~~-2\eta\Delta^{\alpha\nu}\nabla^{\mu}\left(\sigma_{\mu\nu}-\tau_{\pi}\left[\Delta_{\mu}^{\gamma}\Delta_{\nu}^{\rho} D \sigma_{\gamma\rho} + \frac{1}{3}\sigma_{\mu\nu} \theta \right]\right).
\label{eu1}
\end{aligned}
\end{equation}

The CKCJ solutions~\cite{Csorgo:2018pxh} and perturbation solutions here are both characterized by the flow velocity field Eq.~(\ref{e:velocity}) in
the Rindler coordinates. It is straightforward to find that comoving derivative $D$ and expansion rate $\theta$ can be expressed as~\cite{She:2019wdt},
\begin{eqnarray}
D&=&{\rm cosh}(\Omega-\eta_{s})\frac{\partial}{\partial\tau}+\frac{1}{\tau}{\rm sinh}(\Omega-\eta_{s})\frac{\partial}{\partial\eta_{s}},
\label{9}
\end{eqnarray}
and
\begin{eqnarray}
\theta&=&{\rm sinh}(\Omega-\eta_{s})\frac{\partial\Omega}{\partial\tau}+\frac{1}{\tau}{\rm cosh}(\Omega-\eta_{s})\frac{\partial\Omega}{\partial\eta_{s}},
\label{10}
\end{eqnarray}
respectively.

With the help of the Gibbs thermodynamic relation and CNC solutions~\cite{Csorgo:2006ax}, for systems without bulk viscosity and net charge current
(net baryon, net electric charge or net strangeness), the  hydrodynamic conservation equations Eqs.~(\ref{en1},\ref{eu1}) for a longitudinal expanding fireball in the presence of shear viscosity
in the Rindler coordinate can be written as,
\begin{equation}
\begin{aligned}
\tau\frac{\partial T}{\partial\tau}+\tanh(\Omega-\eta_{s})\frac{\partial T}{\partial\eta_{s}}+\frac{\Omega'}{\kappa}T&=\frac{\Pi_{d}}{\kappa}\frac{\Omega'^{2}}{\tau}\cosh(\Omega-\eta_{s})-\frac{\Pi_{d}\tau_{\pi}}{6\kappa\tau^{2}}\Omega'[-6\cosh(2(\Omega-\eta_{s}))\Omega'\\
&~~~~+(1+7\cosh(2(\Omega-\eta_{s})))\Omega'^{2}+
\sinh(2(\Omega-\eta_{s}))\Omega''],
\label{enro}
\end{aligned}
\end{equation}
and
\begin{equation}
\begin{aligned}
\tanh(\Omega-\eta_{s})\left[\tau\frac{\partial T}{\partial\tau}+T\Omega'\right]+\frac{\partial T}{\partial\eta_{s}}&=
\frac{\Pi_{d}}{\tau}(2\Omega'(\Omega'-1)+\Omega''\coth(\Omega-\eta_{s}))\sinh(\Omega-\eta_{s})\\
&~~~~+\frac{\Pi_{d}\tau_{\pi}}{6\kappa\tau^{2}}(-\tanh(\Omega-\eta_{s})\Omega'(12+24\cosh(2(\Omega-\eta_{s}))\\
&~~~~+\Omega'(-28
-46\cosh(2(\Omega-\eta_{s}))+3(5+7\cosh(2(\Omega-\eta_{s})))\Omega'))\\
&~~~~+(18\cosh(2(\Omega-\eta_{s}))+(1-23\cosh(2(\Omega-\eta_{s})))\Omega')\Omega''\\
&~~~~-6\Omega''-3\sinh(2(\Omega-\eta_{s}))\Omega^{(3)}),
\label{euro}
\end{aligned}
\end{equation}
where $\Pi_{d}=\frac{4\eta}{3s}$ is related to the shear viscosity ratio, $\Omega'$ approximately characterizes the longitudinal acceleration of flow element in the medium,  $\Omega',~\Omega'',~\Omega^{(3)}$ are derivative function of flow rapidity, $\Omega'=\frac{\partial \Omega}{\partial \eta_{s}}$, $\Omega''=\frac{\partial^{2}\Omega}{\partial \eta_{s}^{2}}$, $\Omega^{(3)}=\frac{\partial^{3}\Omega}{\partial \eta_{s}^{3}}$.

\emph{\textbf{\underline{Case A}}. Bjorken solution, CNC solution, CKCJ solution under the velocity field $\Omega=\lambda\eta_{s}$.}

A comprehensive study of the longitudinally expanding fireballs for ideal hydrodynamics has been undertaken by Cs\"org\H{o},~Nagy, and~Csan\'ad ( or CNC family of solutions) in Refs.~\cite{Csorgo:2006ax,Csorgo:2008prc}.
According to the results from CNC solutions, the fluid rapidity $\Omega(\tau, \eta_{s})$ depends on space-time rapidity $\eta_{s}$ alone.
For ideal hydrodynamics, one finds $\Omega''=0$, $\Omega^{(3)}=0$, $\Pi_{d}= 0$ and $\tau_{\pi}=0$, the conservation equations Eqs.~(\ref{enro}, \ref{euro}) reduce to,

\begin{equation}
\begin{aligned}
\tau\frac{\partial T}{\partial\tau}+\tanh(\Omega-\eta_{s})\frac{\partial T}{\partial\eta_{s}}+\frac{\Omega'}{\kappa}T=0,
\label{enr}
\end{aligned}
\end{equation}
and
\begin{equation}
\begin{aligned}
\tanh(\Omega-\eta_{s})\left[\tau\frac{\partial T}{\partial\tau}+T\Omega'\right]+\frac{\partial T}{\partial\eta_{s}}=0.
\label{eur}
\end{aligned}
\end{equation}

(a) For boost-invariant Hwa-Bjorken flow where $\Omega = \eta_s$, $\tanh(\Omega-\eta_{s})=0$, and Eqs.~(\ref{enr}, \ref{eur}) have following exact solution,
\begin{equation}
\begin{aligned}
T(\tau) = T_{0}\left(\frac{\tau_{0}}{\tau}\right)^{\frac{1}{\kappa}},
\label{BjorkenT}
\end{aligned}
\end{equation}
where $T_{0}$ define the values for temperature at the proper time $\tau_{0}$ and coordinate rapidity $\eta_{s}=0$.
This is the famous boost-invariant Hwa-Bjorken solution~\cite{Hwa:1974gn, Bjorken:1982qr} and other hydrodynamics variables are functions of the proper time $\tau$.

(b) For a perfect fluid with longitudinal acceleration, $\Omega(\eta_{s})$, shear viscosity $\Pi_{d}= 0$. Eqs.~(\ref{enr}, \ref{eur}) reduce to ,
\begin{equation}
\begin{aligned}
& \tanh(\Omega-\eta_{s})\left[\frac{\tau\partial T}{T\partial\tau}+\Omega'\right]+\frac{\partial T}{T\partial\eta_{s}}=0,\\
\Rightarrow ~~~~& \frac{\partial\ln T}{\partial\ln\tau}=-\frac{1-\kappa\tanh^{2}(\Omega-\eta_{s})}{1-\tanh^{2}(\Omega-\eta_{s})}\frac{\Omega'}{\kappa}.
\label{cnc1}
\end{aligned}
\end{equation}
For {\bf\large $\kappa=1$}, one finds that the $\Omega'$ could be either an arbitrary constant~\cite{Csorgo:2006ax} or a function of $\eta_{s}$~\cite{Csorgo:2018pxh,Kasza:2018qah}. For the former case, one finds that $\Omega(\eta_{s})=\lambda\eta_{s}+\eta_{0}$, here $\lambda$ is an arbitrary constant and $\lambda=\Omega'$, $\eta_{0}$ is a space-time rapidity shift. The solution of Eq.~(\ref{cnc1}) case is
\begin{equation}
\begin{aligned}
\frac{\partial\ln T}{\partial\ln\tau}=-\Omega',
~~~\Rightarrow~~~T(\tau)=T_{0}\left(\frac{\tau_{0}}{\tau} \right)^{\lambda}\frac{1}{\mathcal V(S)},
\label{cncs}
\end{aligned}
\end{equation}
this is the well-known CNC exact solutions case (e) that presented by Cs\"org\H{o},~Nagy, and~Csan\'ad in Refs.~\cite{Csorgo:2006ax,Csorgo:2008prc}. Here $\mathcal V(S)$ is an arbitrary function of scaling function $S$, where $S$ can be obtained from the scaling function definition $u^{\mu}\partial_{\mu}S=0$, the properties and detailed discussion of this $S$ can be found in Ref.~\cite{Csorgo:2008prc} for CNC exact solutions and in Ref.~\cite{Csorgo:2020iug} for Hubble-type viscous flow.

(c) A finite and accelerating, realistic 1+1 dimensional solution of relativistic hydrodynamics was recently given by Cs\"org\H{o}, Kasza, ~Csan\'ad and Jiang (CKCJ)
with the condition $H(\eta_{s})=\Omega(\eta_{s})-\eta_{s}$,
for details, see review in Ref.~\cite{Csorgo:2018pxh,Kasza:2018qah}.

\emph{\textbf{\underline{Case B}}. Perturbation solution with Navier-Stokes approximation.}

For a relativistic hydrodynamic in the Navier-Stokes (first order) approximation, fluid flow rapidity $\Omega=\lambda\eta_{s}=(1+\lambda^{*})\eta_{s}$, shear viscosity tensor $\pi^{\mu\nu}=2\eta\sigma^{\mu\nu}$, shear viscosity ratio $\Pi_{d}=4\eta/3s$, the relaxation time $\tau_{\pi}=0$. The last terms in the right of Eqs.~(\ref{enro},~\ref{euro}) disappear automatically as follow,

\begin{equation}
\begin{aligned}
\tau\frac{\partial T}{\partial\tau}+\tanh(\lambda^{*}\eta_{s})\frac{\partial T}{\partial\eta_{s}}+\frac{1+\lambda^{*}}{\kappa}T&=\frac{\Pi_{d}}{\kappa}\frac{1+2\lambda^{*}+\lambda^{*2}}{\tau}\cosh(\lambda^{*}\eta_{s}),
\label{enrn}
\end{aligned}
\end{equation}
and
\begin{equation}
\begin{aligned}
\tanh(\lambda^{*}\eta_{s})\left[\tau\frac{\partial T}{\partial\tau}+T(1+\lambda^{*})\right]+\frac{\partial T}{\partial\eta_{s}}&=
\frac{\Pi_{d}}{\tau}(2(1+\lambda^{*})\lambda^{*})\sinh(\lambda^{*}\eta_{s}).
\label{eurn}
\end{aligned}
\end{equation}

However, the reduced conservation equations Eqs.~(\ref{enrn},~\ref{eurn}) including first order approximation are still a set of nonlinear differential equations, which are notoriously hard to solve analytically. Fortunately, based on the results from the ideal hydro~\cite{Jiang2017,Jiang:2018qxd}, we found that the longitudinal acceleration parameter $\lambda^{*}$ extracted from the experimental data is pretty small ($0<\lambda^{*}\ll1$), which resulting in a simply antsaz or perturbation solution here.

We assume the longitudinal rapidity perturbation $\lambda^{*}\eta_{s}$ is a pretty small numbers here. Using the Taylor series expansion $\tanh(\lambda^{*}\eta_{s})\approx\lambda^{*}\eta_{s}-\frac{(\lambda^{*}\eta_{s})^{3}}{3}$, $\cosh(\lambda^{*}\eta_{s})\approx1+\frac{(\lambda^{*}\eta_{s})^{2}}{2}$ and $\sinh(\lambda^{*}\eta_{s})\approx\lambda^{*}\eta_{s}+\frac{(\lambda^{*}\eta_{s})^{3}}{6}$, up to the leading order ${\mathcal O}(\lambda^{*})$, Eqs.~(\ref{enrn},~\ref{eurn}) yields a partial differential equation depending on $\tau$ only,
\begin{equation}
\begin{aligned}
\tau\frac{\partial T}{\partial\tau}+\frac{(1+\lambda^{*})T}{\kappa}=\frac{\Pi_{d}}{\kappa}\frac{1+2\lambda^{*}}{\tau},
\label{enrnt}
\end{aligned}
\end{equation}
and the exact temperature solution $T(\tau,\eta_{s})$ of above equation is
\begin{equation}
\begin{aligned}
T(\tau,\eta_{s})&=T_{1}(\eta_{s})\left(\frac{\tau_{0}}{\tau}\right)^{\frac{1+\lambda^{*}}{\kappa}}+\frac{(2\lambda^{*}+1)\Pi_{d}}{(\kappa-1)\tau_{0}}\left(\frac{\tau_{0}}{\tau}\right)^{\frac{1+\lambda^{*}}{\kappa}}
\left[1-\left(\frac{\tau_{0}}{\tau}\right)^{1-\frac{1+\lambda^{*}}{\kappa}}\right],
\label{tt1}
\end{aligned}
\end{equation}
where $\tau_{0}$ is the value of proper time, $T_{1}(\eta_{s})$ is an unknown function constrained by Eq.~(\ref{eurn}).

Putting Eq.~(\ref{tt1}) back to the Euler equation Eq.~(\ref{eurn}), up to the leading order ${\mathcal O}(\lambda^{*})$, one gets the exact expression of $T_{1}(\eta_{s})$ as follow, \\
\begin{equation}
\begin{aligned}
T_{1}(\eta_{s})&=T_{0}\exp[-\frac{1}{2}\lambda^{*}(1-\frac{1}{\kappa})\eta_{s}^{2}]
-\frac{\left(1-\exp[-\frac{1}{2}\lambda^{*}(1-\frac{1}{\kappa})\eta_{s}^{2}] \right)\Pi_{d}}{(\kappa-1)\tau_{0}},
\label{t1}
\end{aligned}
\end{equation}
where $T_{0}$ define the values for temperature at the proper time $\tau_{0}$ and coordinate rapidity $\eta_{s}=0$.
Here if one puts the perturbation solution Eq. (22) back to the energy conservation equation Eq. (19),  up to the leading order ${\mathcal O}(\lambda^{*})$, one can obtain the same results.

Finally, inputting Eq.~(\ref{t1}) into Eq.~(\ref{tt1}), the perturbation solution of the $1+1$ D embeding $1+3$ D relativistic viscous hydrodynamics can be written as,\\
\begin{equation}
\begin{aligned}
T(\tau,\eta_{s})&=T_{0}\left(\frac{\tau_{0}}{\tau}\right)^{\frac{1+\lambda^{*}}{\kappa}}
{\bigg [}\exp(-\frac{1}{2}\lambda^{*}(1-\frac{1}{\kappa})\eta_{s}^{2})+\frac{R_{0}^{-1}}{\kappa-1}{\bigg (}2\lambda^{*}+\exp[-\frac{1}{2}\lambda^{*}(1-\frac{1}{\kappa})\eta_{s}^{2}]-(2\lambda^{*}+1)
\left(\frac{\tau_{0}}{\tau}\right)^{\frac{\kappa-\lambda^{*}-1}{\kappa}} {\bigg )}
 {\bigg ]},
\label{t2}
\end{aligned}
\end{equation}
where the Reynolds number is $R^{-1}_{0}=\frac{\Pi_{d}}{T_{0}\tau_{0}}$~\cite{AM:2004prc,Kouno:1989ps}.

From a perspective of the perturbation solution's structure, one finds above conditional perturbation solution is very nontrivial since it only satisfied when the longitudinal accelerating parameter $\lambda^{*}\ll1$ and $\eta_{s}$ is not very large ($|\eta_{s}|\ll 5$), furthermore, it involves two different transport coefficients and many nonvanishing components of the longitudinal expanding properties.
In order to investigate the stability of perturbation solution Eq.~(\ref{t2}),
we furthermore numerically solved the energy equation Eq.~(\ref{enrn}) and Euler equation Eq.~(\ref{eurn}) with conditions $\Pi_{d}=0$ and $\tau_{\pi}=0$, the longitudinal accelerating parameter $\lambda^{*}=0.05$, the grid length of proper time $\Delta\tau=0.05$, the grid of space-time rapidity $\Delta\eta_{s}=0.05$, and the range space-time rapidity $\eta_{s}$ from 0.0 to 5.0.
The comparison of the perturbation solution and the numerical solution are presented in Fig.\ref{fig1} left panel, the difference between above two solutions appears to be small, which means the perturbation solution is a special but stable one.

The profile of Eq.~(\ref{t2}) is a (1+1) dimensional scaling solution in (1+3)
dimensions and the $\eta_{s}$ dependence of temperature density is of the Gaussian form, see Fig. \ref{fig1} right panel.
Such perturbation solutions implies that for a non-vanishing longitudinal acceleration parameter $\lambda^{*}$, the cooling rate is larger than for the ideal case.
Meanwhile, a non-zero shear viscosity $\eta$ makes the cooling rate smaller than for the ideal case~\cite{Jiang:2018qxd}, see Fig. \ref{fig1} left panel.
Note that when $\lambda^{*}=0$ and $R_{0}^{-1}=0$, one obtains the same solutions as same as the ideal hydrodynamic Bjorken solution~\cite{Bjorken:1982qr}, when $\lambda^{*}=0$ and $R_{0}^{-1}\neq0$, one obtains the first order Bjorken solutions~\cite{AM:2004prc,DTeaney}, if $\lambda^{*}\neq0$ and $R_{0}^{-1}=0$, one obtains a special solution which is consistent with the CNC solutions' case (e) in~\cite{Csorgo:2006ax,Csorgo:2008prc}, and
when one solve the Eqs.~(\ref{enr},~\ref{eur}) directly with $R_{0}^{-1}=0$ and $\lambda^{*}\neq0$, one obtains the CKCJ solutions~\cite{Csorgo:2018pxh}.

\emph{\textbf{\underline{Case C}}. Perturbation equations with Israel-Stewart approximation.}

The temperature profile Eq.~(\ref{t2}) shows a peak at earlier proper time $\tau$ in the Navier Stokes approximation, see Fig.\ref{fig1}.
The source of this acausality can be understood from the constitutive relations satisfied by the dissipative currents $\pi^{\mu\nu}=2\eta\sigma^{\mu\nu}$.
The linear relationship between dissipative currents and gradients of the primary fluid-dynamical variables imply that any inhomogenity of $u^{\mu}$, immediately
results in dissipative currents. This instantaneous effect causes the first order theory to be unstable at earlier times.

Fortunately, people found that the Israel-Stewart (second order) approximation are suitable in describing the physical process happening at earlier times, and it describes the counteract of acceleration effect and viscosity effect well.
However, it's hard to solve the the differential equations Eqs.~(\ref{enro},~\ref{euro}) analytically with the Israel-Stewart approximation.
So we numerically solve the temperature time dependence Eq.~(\ref{enro}) first at $\eta_{s}=0.0$ with the initial condition $T(0.2, 0.0)=0.65$ GeV first, here the grid length of $\tau$ is $\Delta\tau=0.05$ fm. Then, for each $\eta_{s}$, we solve the temperature rapidity dependence Eq.~(\ref{euro}) step by step with the results from the Eq.~(\ref{enro}), and solve these equations together, the grid length of $\tau$ is $\Delta\tau=0.05$, too.
The temperature distribution of thermodynamic quantities ($\varepsilon,~T,~p$) in whole $(\tau,~\eta_{s})$ coordinates with initial condition $T(\tau_{0},~\eta_{s0})$ now is a Gaussian shape, see Fig.~\ref{fig2}.
Furthermore, in order to compare with the perturbation results from the first order approximation, the Eqs.~(\ref{enro},~\ref{euro}) can be rewritten up to the leading order ${\mathcal O}(\lambda^{*})$ as follow,
\begin{eqnarray}
\tau\frac{\partial T}{\partial\tau}&=&\frac{(2\lambda^{*}+1)\Pi_{d}}{3\tau}-\frac{(\lambda^{*}+1)T}{3}-\frac{(2-\ln 2)\Pi_{d}(1+6\lambda^{*})}
{9 \pi T \tau^{2}}, \label{en2s}\\
\frac{\partial T}{\partial\eta_{s}}&=&\lambda^{*}\eta_{s}\left[-\frac{2T}{3}-\frac{\Pi_{d}}{3\tau}+\frac{(2-\ln 2)\Pi_{d}}{9\pi T \tau^{2}} \right]. \label{eu2s}
\end{eqnarray}
Above differential equations~(\ref{en2s},~\ref{eu2s}) can not be solved analytically, we using the same numerically method as for the Eqs.~(\ref{enr},~\ref{eur}), we solve the above second-order viscous hydrodynamic equations~(\ref{en2s},~\ref{eu2s}) with the conformal equation of state $\varepsilon~=~3p$ and relaxation time $\tau_{\pi}=\frac{2-\ln2}{2\pi T}$~\cite{Policastro:2001yc,Baier:2007ix,Bhattacharyya:2008jc,Arnold:2011ja} directly in the Rindler coordinates, the numerical results are presented in Fig.~\ref{fig3} .

\section{Results and discussion}
\label{section3}

The temperature profiles obtained in the previous section are now applied to study the longitudinal expanding dynamics,
the initial condition $T(\tau_{0},~\eta_{s0})$ can be arbitrarily chosen.
Following the result from~\cite{AM:2004prc}, the initial proper time $\tau_{0}=0.2$ fm/$c$, and initial temperature $T_{0}(0.2,~0.0)=0.65$ GeV are used in the calculation.

\begin{figure}[!ht]
\begin{center}
\includegraphics[trim=0.0cm 0.1cm 0cm 0cm, clip,width=6.8cm, height=5.8cm]{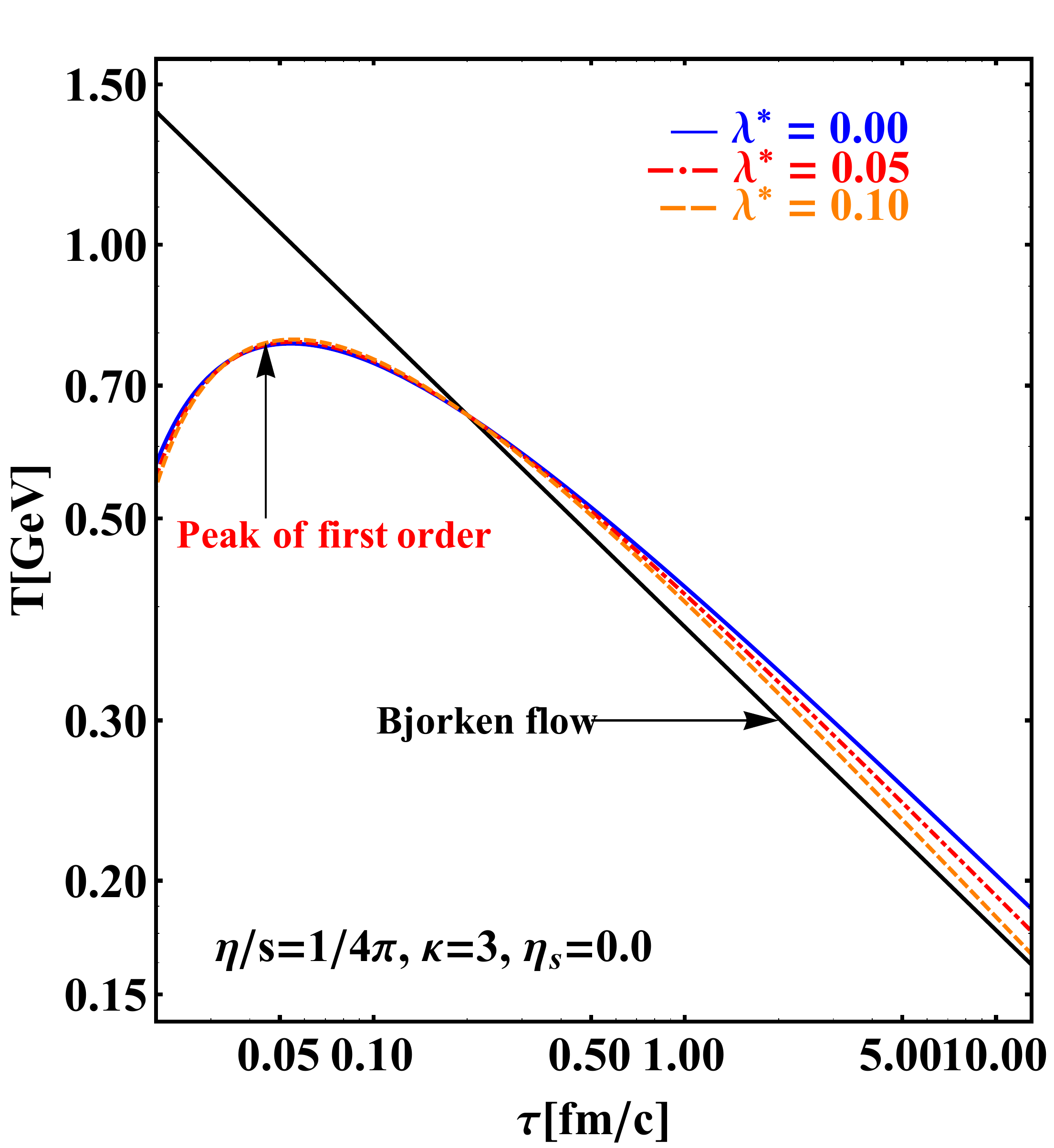}~~~~~~~~~~~~~~~~~
\includegraphics[trim=0.0cm 0.0cm 0cm 0cm, clip,width=6.8cm, height=5.43cm]{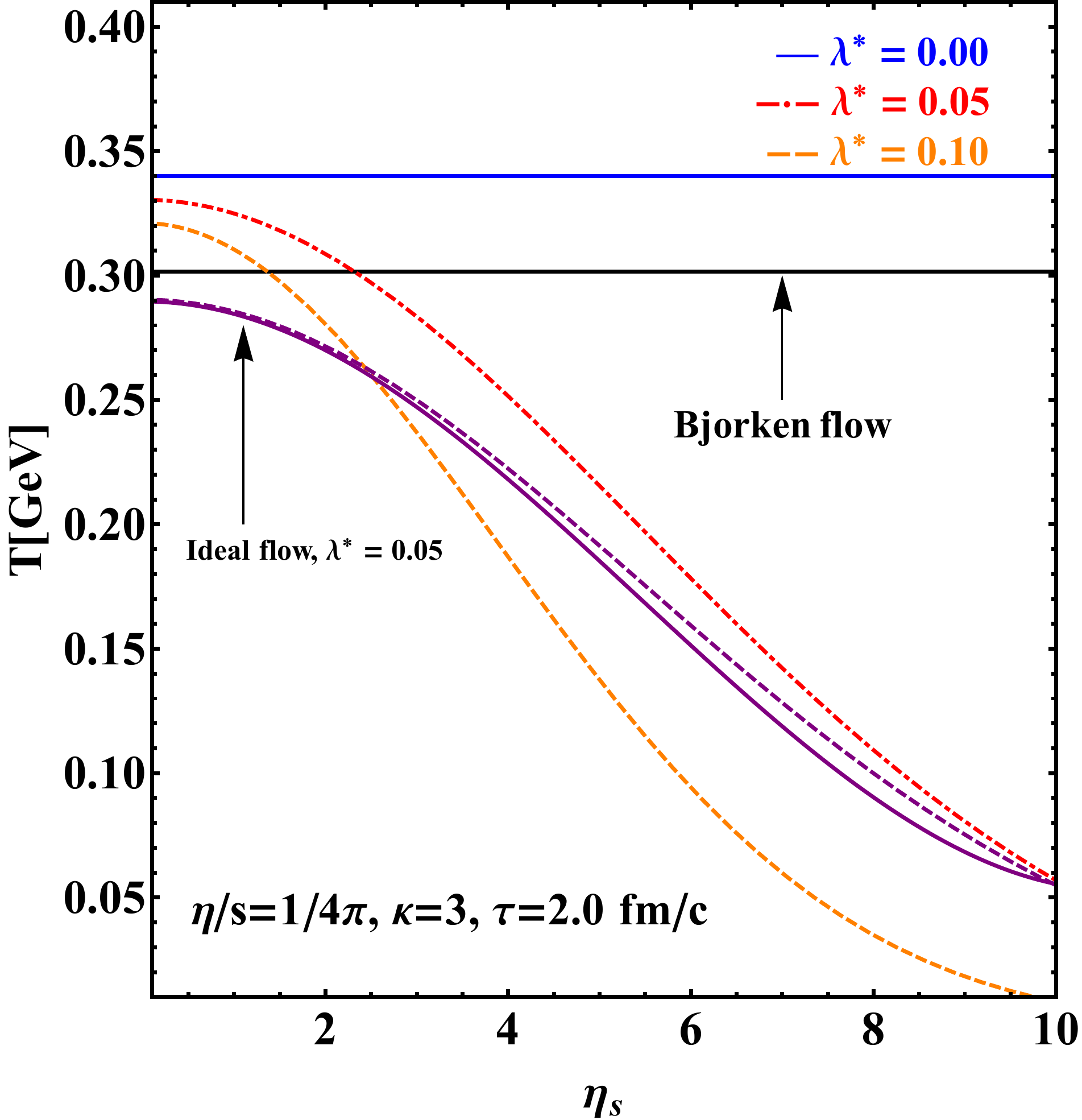}
\end{center}
\caption{(Color online) Temperature profile in the Navier-Stokes approximation for different longitudinal acceleration parameters $\lambda^{*}$.
Equation of state $\varepsilon=3p$, shear viscosity ratio $\eta/s=1/4\pi$. Black solid curve is the ideal Bjorken flow for reference, blue solid curve is the 1st order Bjorken flow. Left panel: The proper time $\tau$ evolution of temperature for $\eta_{s}=0$. Right panel: The space rapidity $\eta_{s}$ evolution of temperature for $\tau= 2$ fm/$c$. Purple curve show the comparison between perturbation solution (dashed) and numerical solution (solid) for ideal flow with $\lambda^{*}=0.05$, the accuracy is acceptable in the range $0.0\leq\eta_{s}\leq5.0$. Results from the perturbation solution Eq.(\ref{t2}) and numerical solution for Eqs.~(\ref{enro},\ref{euro}).}
\label{fig1}
\end{figure}

Fig.~\ref{fig1} show the longitudinal expanding effect dependence of temperature evolution in the Navier-Stokes approximation.
In the left panel of Fig.~\ref{fig1} shows the time-dependence of the temperature for different viscosity and the longitudinal acceleration parameter $\lambda^{*}$. The black curve is the ideal Bjorken flow. It is seen that the larger the longitudinal acceleration parameter $\lambda^{*}$, the faster the medium cool down. However, the viscosity effect slow down the medium cooling. It is important to note that there is a peak at early time in $T$ in the case of first order approximation.
In the right panel of Fig.~\ref{fig1} shows the space-time rapidity dependence of the temperature at $\tau=$ 2 fm/$c$. The temperature distribution of ideal Bjorken flow (black curve) and the Bjorken flow under Navier-Stokes limit (blue curve) show a flat-plateau shape. The effect of the longitudinal accelerating expanding, however, make temperature distribution to a Gaussian shape (red and orange curve). In addition, the difference between the numerical solution (purple solid curve) and the perturbation solution
(purple dashed curve) for $\Pi_{d}=\tau_{\pi}=0$ are presented, one finds that the difference in the range $0.0\leq\eta_{s}\leq5.0$ is acceptable.

\begin{figure}[!ht]
\begin{center}
\includegraphics[trim=0.0cm 0.1cm 0cm 0cm, clip,width=6.8cm, height=5.8cm]{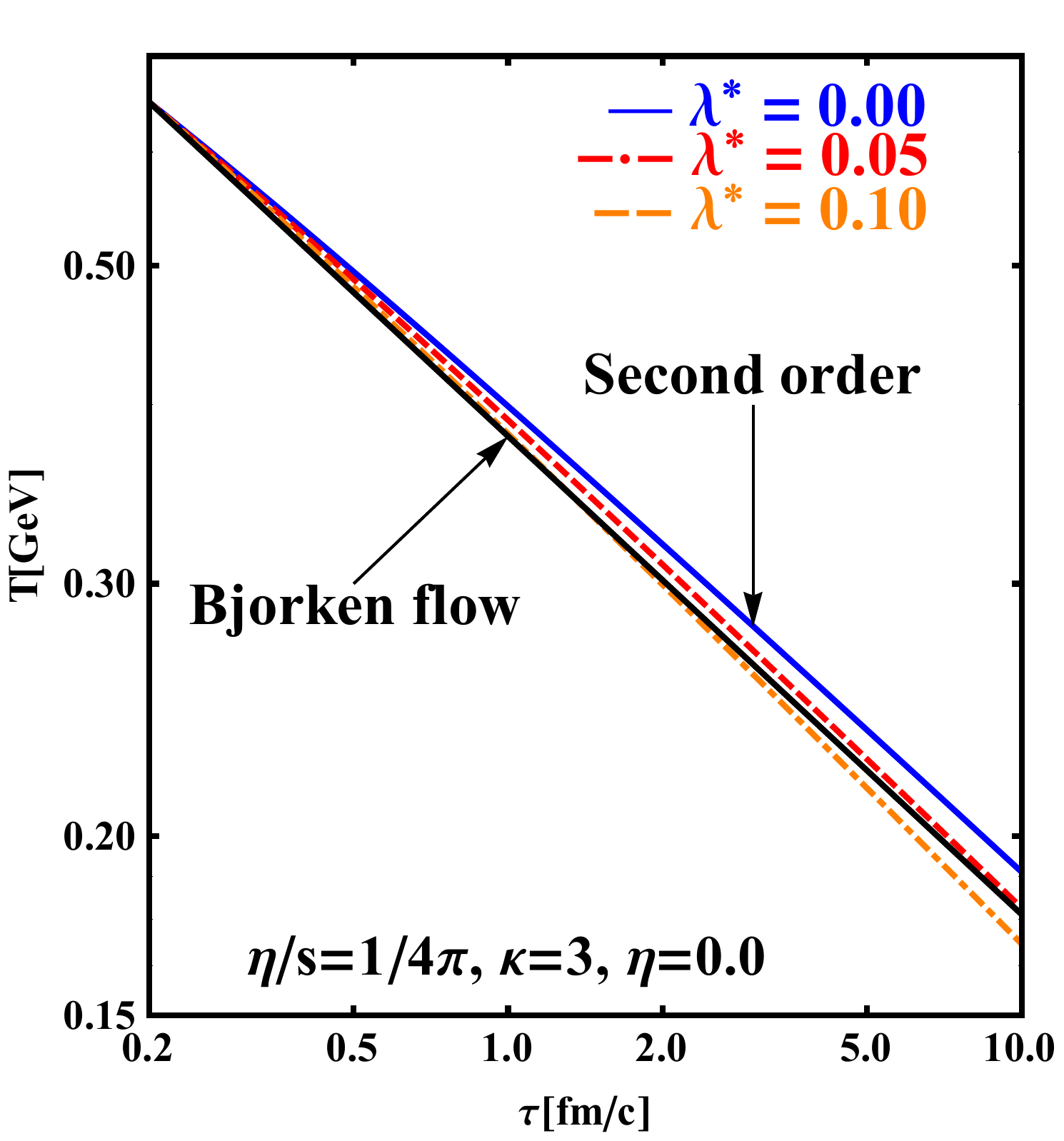}~~~~~~~~~~
\includegraphics[trim=0.0cm 0.0cm 0cm 0cm, clip,width=6.8cm, height=5.43cm]{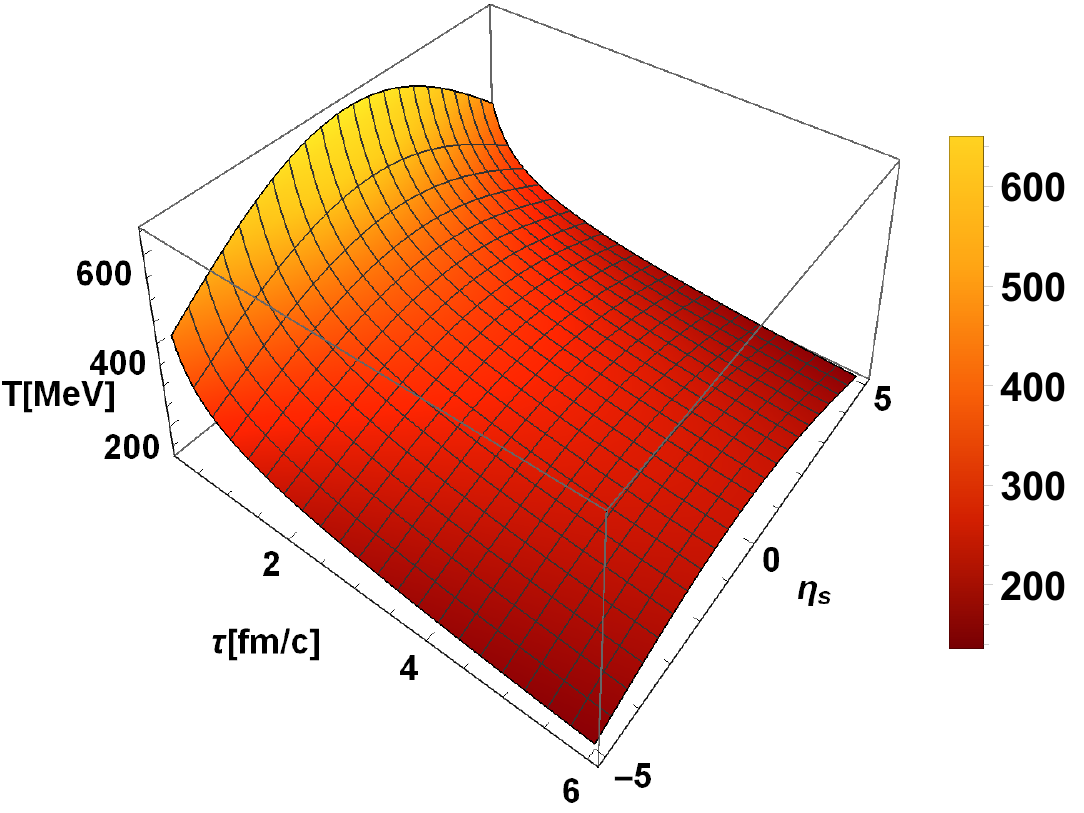}
\end{center}
\caption{(Color online) Temperature profile in the Israel-Stewart approximation. Left panel: The proper time $\tau$ evolution of temperature for $\eta_{s}=0$. Black solid curve is the ideal Bjorken flow for reference. Right panel: The space-time evolution of temperature in ($\tau-\eta_{s}$) coordinates, the longitudinal acceleration parameter $\lambda^{*}=0.1$. Numerical results from Eqs. (\ref{enr},~\ref{eur}).}
\label{fig2}
\end{figure}

Fig.~\ref{fig2} show the completely temperature evolution for different longitudinal acceleration parameter $\lambda^{*}$ in the Israel-Stewart approximation.
In the left panel of Fig.~\ref{fig2} shows the time-dependence of the temperature, one finds no peak at the early time of $T$, the first order theory significantly underpredicts the work done during the expansion relative to the Israel-Stewart approximation. One also finds the effect of viscous compensates the effect from longitudinal acceleration when $\eta/s=1/4\pi$ and $\lambda^{*}=0.05$ at larger proper time of evolution, the viscous curve (red dashed) almost overlaps with the Bjorken flow (black solid).
The longitudinal expanding effect make the medium cool down fast and there is no peak at early time in $T$.
In the right panel of Fig.~\ref{fig2} shows the temperature distribution in ($\tau,\eta_{s}$) coordinates with $\lambda^{*}=0.1$.

\begin{figure}[!ht]
\begin{center}
\includegraphics[trim=0.0cm 0.1cm 0cm 0cm, clip,width=6.8cm, height=5.8cm]{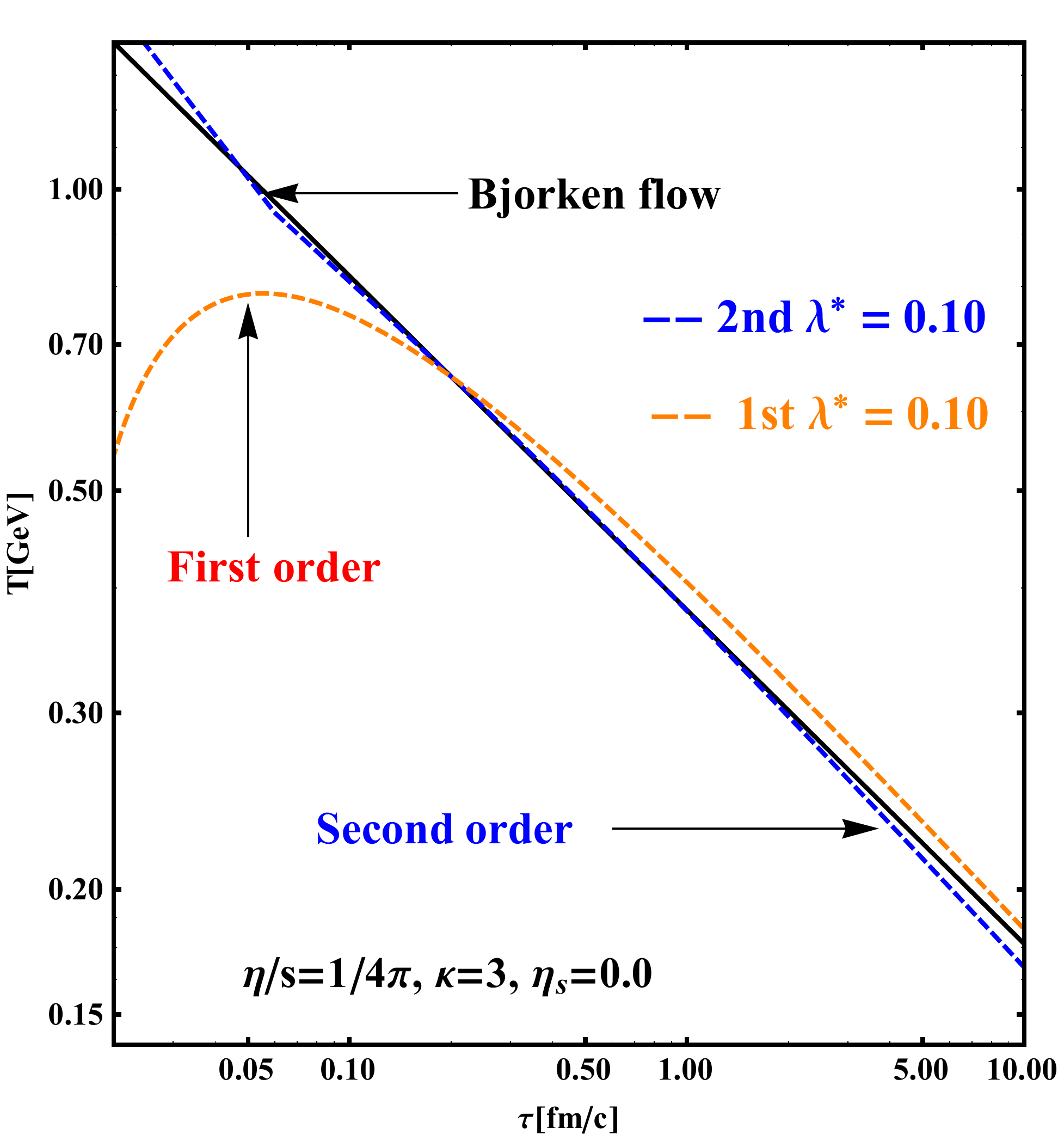}~~~~~~~~~~
\includegraphics[trim=0.0cm 0.1cm 0cm 0cm, clip,width=6.8cm, height=5.8cm]{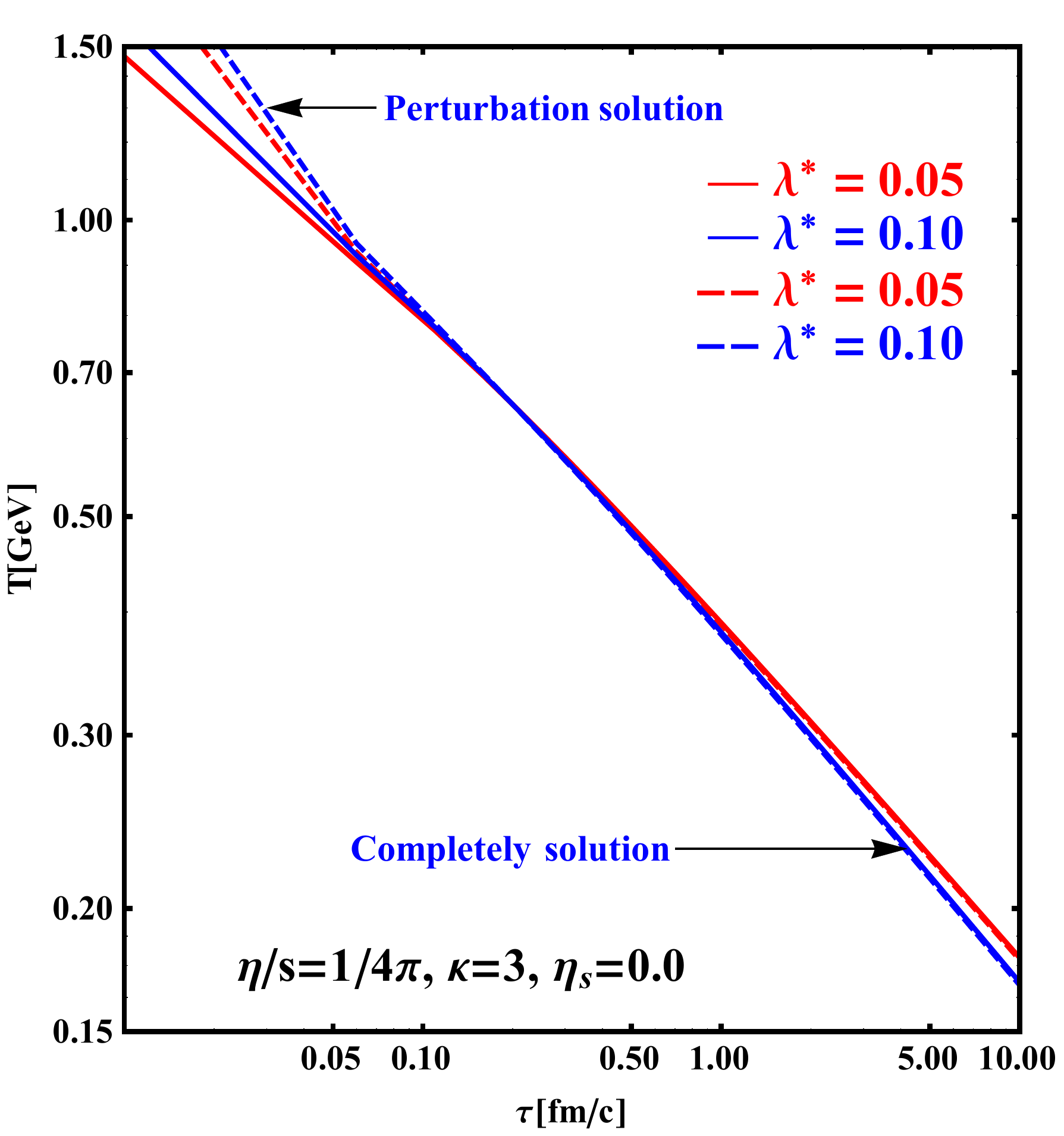}
\end{center}
\caption{(Color online) The proper time evolution of temperature density for given primary initial conditions. Left panel: Perturbation results of temperature profile in the Navier-Stokes approximation (1st) and in the Israel-Stewart theory (2nd). The longitudinal acceleration parameter $\lambda^{*}=0.10$. Black solid curve is the ideal Bjorken flow for reference. Right panel: Temperature profile comparsion between the completely solution (solid curve) with perturbation solutions (dashed curve) in the Israel-Stewart theory for different $\lambda^{*}$, the grid of $\tau$ is $\Delta\tau=0.05$ in the numerical code.}
\label{fig3}
\end{figure}

So far our focus has been on study the temperature evolution of perturbation solutions through the Navier-Stokes theories and Israel-Stewart theories independently. Now we analyze
the difference between these two theories under the same longitudinal acceleration effect. We numerically solve the differential equations Eqs.~(\ref{en2s},~\ref{eu2s}) together with the initial condition $T_{0}(0.2,~0.0)=0.65$ GeV first. In the left panel of Fig.~\ref{fig3} show the comparison of the second order perturbation solutions, the first order perturbation solutions and Bjorken solution. In the right panel of Fig.~\ref{fig3} show the comparison between the second order perturbation solutions and the completely numerical results. For small $\lambda^{*}$, we find that the perturbation solutions are stable and show good agreement with completely numerical results at large time.

\section{Summary}
\label{section4}

We have investigated the relativistic viscous hydrodynamics for longitudinal expanding fireballs
in terms of the Navier-Stokes theory and Israel-Stewart theory by embedding 1+1 D fluid into a 1+3 D space-time.
The results obtained in this paper are summarized as follows.

(1) We expand the current knowledge of accelerating hydrodynamics~\cite{Csorgo:2006ax,Csorgo:2008prc,Csorgo:2018pxh} by including the second-order viscous corrections in the relativistic hydrodynamics fluid with
longitudinal expanding fireballs and general equation of state. The effect of longitudinal acceleration accelerates the thermodynamics evolution of medium while the viscosity effect decelerates the evolution in the Minkowski space-time.

(2) The perturbation solution from the Navier-Stokes approximation is explicit and simple in mathematical structure, and it is consistent with the results from Refs.\cite{Jiang:2018qxd}. Furthermore, the comparison between the perturbation solution and the full numerical solution are investigated, as we presented in Fig.~\ref{fig1} right panel, and it shows that perturbation approximation for $\lambda\eta_{s}$ are valid in the leading order accuracy of the longitudinal acceleration parameter $\lambda^{*}$. The temperature distribution here indicates a Gaussian shape in the $\eta_{s}$ direction.

(3) For small perturbations along the longitudinal directions, as we presented in Fig.~\ref{fig1} right panel, the perturbation solution from the Navier-Stokes approximation is stable in region $\tau\geq\tau_{0}$ of while it is unstable in region $\tau\leq\tau_{0}$.

(4) The numerical results from the Israel-Stewart approximation in longitudinal expansion relativistic viscous hydrodynamics solve the causal problem and the temperature profile
in the Rindler coordinate are presented.

There are still many open questions about such perturbation solutions and results.

(1) For consistency and stability, the perturbation solution is meaningful when $\lambda^{*}$ is pretty small and $|\lambda^{*}\eta_{s}|\ll1$,
for arbitrary longitudinal acceleration parameter $\lambda^{*}$, e.g. $|\lambda^{*}|\gg 1$,
such perturbation approximation become unsuitable and we need to solve the differential equations completely by other numerical method, such as 3+1D CLvisc~\cite{Pang:2018zzo}.
In addition, if one treats the fluid rapidity $\Omega$ as an unknown function of both the proper time $\tau$ and space-time rapidity $\eta_{s}$, the conservation equations will be extremely complicated than current Eqs.~(\ref{enro},\ref{euro}) even for the Navier-Stokes approximation, and the analytical solution is hard to get, for more discussion about this issue, see~\cite{Csorgo:2008prc}.
(2) To find new exact solution of hydrodynamics, it is possible to use the method from the AdS/CFT theory~\cite{Janik:2005zt,Janik:2006ft}, which provides a method that search the exact solutions by expanding in the small and large proper time $\tau$ limit.
(3) The shear pressure tensor relaxation time $\tau_{\pi}$ assumed above for the second theory is definitely oversimplified, and it is only valid for smaller values of     $\tau_{\pi}$. While physically motivated, we acknowledge that this method is imperfect.
(4) Transverse expansion cannot be neglected, especially during the later stages of the fireball,
significantly changing the observables at RHIC and LHC. In reality the expansion of the
system will not be purely longitudinal, the system will also expand transversally~\cite{Gubser:2010ui,Hatta:2014gqa}.
(5) It is important to note that the QGP bulk viscosity ratio $\zeta/s$ is not zero from the lattice QCD calculation,
the effect of bulk viscosity property play a curial role when temperature is larger than 3$T_{c}$~\cite{Meyer:2007dy,Kharzeev:2007wb}.
Recently, new solutions of first order viscous hydrodynamics for Hubble-type flow are presented to study the bulk viscosity~\cite{Csanad:2019lcl}, however,
the second order theory of such fluid is still unknown.
(6) In principle, the second order approximation should depend on a larger number of independent transport coefficient,
e.g. $\eta$, $\tau_{\pi}$, $\lambda_{1}$, $\lambda_{2}$, $\lambda_{3}$, and,
the direction extension results Eqs.~(\ref{en1},~\ref{eu1}) from the $\partial_{\mu}s^{\mu}\geq 0$ are, in fact, incomplete and ad-hoc.
In order to determine these transport coefficients, microscopic theories, such as kinetic theory should be studied~\cite{Baier:2006um}.
(7) Chapman-Enskog expansion and completely Grid's 14-moment methods~\cite{Bhalerao:2013pza} could be used to study the higher order correction.
(8) Recently, Duke group presented a novel method to study the effective viscosities~\cite{Paquet:2019npk}, which points a new way to study the shear viscosity and bulk viscosity for QGP.
As a next step, we try to study above parts in more accurate studies in the future.

\begin{acknowledgements}
We especially thank Xin-Nian Wang, M.~Csan\'ad, T.~Cs\"org\H{o}, N.~I.~Nagy and Chao Wu for valuable comments at the initial stage of this study.
This work was in part supported by the Ministry of Science and Technology of China (MSTC) under the "973" Project No. 2015CB856904(4), by NSFC Grant Nos. 11735007, 11890711.
This work was supported by the Sino-Hungarian bilateral cooperation program, under the Grand No.Te'T 12CN-1-2012-0016. D. She is supported by the China Scholarship Council (CSC) Contract No.201906770027. Z-F. Jiang would like to thank T.~Cs\"org\H{o}, M.~Csan\'ad, L{\'e}vai P{\'e}ter and Gergely G{\'a}bor Barnafoldi for kind hospitality during his stay at Winger RCP, Budapest, Hungary.

\end{acknowledgements}

\end{document}